\documentclass[11pt,a4paper]{article}
\usepackage[
  twoside,
  paperwidth=210mm,
  paperheight=297mm,
  textheight=562pt,
  textwidth=384pt,
  centering,
  headheight=50pt,
  headsep=12pt,
  footskip=24pt,
  footnotesep=24pt plus 2pt minus 12pt,
]{geometry}

\usepackage[colorlinks=true
,urlcolor=blue
,anchorcolor=blue
,citecolor=blue
,filecolor=blue
,linkcolor=blue
,menucolor=blue
,linktocpage=true
,pdfproducer=medialab
]{hyperref}
\newcommand{\Li}{\mathrm{Li}}
\newcommand{\PI}{\mathrm{PI}}
\newcommand{\D}{\mathrm{d}}
\newcommand{\I}{\mathrm{i}}
\newcommand{\io}{\mathrm{i0}}
\usepackage{amsmath}
\usepackage{cite}
\usepackage{authblk}
\usepackage[super]{nth}
\usepackage{multirow}
\usepackage{stix}               
\usepackage{cancel}
\usepackage[svgnames,x11names]{xcolor}
\usepackage{tikz}
\definecolor{lightgray}{HTML}{D5D5D5}
\definecolor{lightergray}{HTML}{EEEEEE}
\usepackage{listings}
\newcommand\mmaCU{\scalebox{0.8}{
\hspace{-0.1cm}\begin{tikzpicture}[baseline=4pt]
\def\bh{15pt}
\def\wl{21pt}
\def\wr{31pt}
\draw [lightgray,rounded corners =2.4pt, line width=1pt,fill=lightergray] (\wl, \bh) -- (  0,\bh) -- (  0,0) -- (\wl,0);
\draw [lightgray,rounded corners =2.4pt, line width=1pt                 ] (\wl, \bh) -- (\wr,\bh) -- (\wr,0) -- (\wl,0);
\draw [lightgray,                        line width=1pt                 ] (\wl, \bh) --                         (\wl,0);
\node at (11pt,7pt) {\large\textsf{ctrl}};
\node at (26pt,3pt) {\large--};
\end{tikzpicture}
}}
\newcommand\mmaPlus{\scalebox{0.8}{
\hspace{-0.1cm}\begin{tikzpicture}[baseline=4pt]
\draw [lightgray, rounded corners=2.4pt, line width=1pt] (0,0) rectangle (12pt,15pt);
\node at (6pt,6pt) {\footnotesize\textsf{+}};
\end{tikzpicture}\hspace{-0.1cm}
}}
\newcommand\mmaMinus{\scalebox{0.8}{
\hspace{-0.1cm}\begin{tikzpicture}[baseline=4pt]
\draw [lightgray, rounded corners=2.4pt, line width=1pt] (0,0) rectangle (12pt,15pt);
\node at (6pt,6pt) {\footnotesize\textsf{-}};
\end{tikzpicture}\hspace{-0.1cm}
}}

\def\name{\texttt{handyG}}
\def\basegit{https://gitlab.com/mule-tools/handyG}

\author[1]{L. Naterop}
\author[1,2]{A. Signer}
\author[1,2]{Y. Ulrich\thanks{Corresponding author.\\E-mail address:
\href{mailto:yannick.ulrich@psi.ch}{yannick.ulrich@psi.ch}}}

\affil[1]{\sl\small
Physik-Institut, Universit\"at Z\"urich, \protect\\
Winterthurerstrasse 190,
CH-8057 Z\"urich, Switzerland\vspace{0.5cm}
}

\affil[2]{\sl\small Paul Scherrer Institut\protect\\
CH-5232 Villigen PSI, Switzerland}
\date{\vspace{-5ex}}

\title{\name{} -- rapid numerical evaluation of generalised
polylogarithms in Fortran}

\begin{document}
\thispagestyle{empty}
\begin{flushright}
PSI-PR-19-17\\
ZU-TH 40/19\\
\end{flushright}
{\let\newpage\relax\maketitle}
\hrule\vskip12pt
\noindent\unskip\textbf{Abstract}
\par\medskip\noindent\unskip\ignorespaces
Generalised polylogarithms naturally appear in higher-order
calculations of quantum field theories. We present \name{}, a Fortran
90 library for the evaluation of such functions, by implementing the
algorithm proposed by Vollinga and Weinzierl. This allows fast
numerical evaluation of generalised polylogarithms with currently
relevant weights, suitable for Monte Carlo integration.
\par\medskip\noindent\unskip\ignorespaces
\par\vskip10pt
\hrule\vskip12pt

\newpage


{\bf PROGRAM SUMMARY}

\vspace{0.5cm}
\begin{small}
\setlength{\parindent}{0pt}
{\em Program Title:} \name
\bigbreak
{\em Licensing provisions:} GPLv3
\bigbreak
{\em Programming language:}  Fortran 90
\bigbreak
{\em Operating system:}
Linux (tested on Ubuntu 18.04 and Scientific Linux 7.6), macOS. Code
optimisation is only available with recent compilers
\bigbreak
{\em Distribution format:} \url{\basegit}
\bigbreak
{\em E-mail:} \href{mailto:yannick.ulrich@psi.ch}{yannick.ulrich@psi.ch}
\bigbreak
{\em Other programs called:} none, Mathematica interface available
\bigbreak
{\em Nature of problem:}
  Numerical evaluation routine for generalised (or
  Goncharov~\cite{Goncharov}) polylogarithms that is fast
  enough for Monte Carlo integration.
\bigbreak
{\em Solution method:}
  Implementing the algorithm presented by Vollinga and
  Weinzierl~\cite{Vollinga} in Fortran 90, providing a Fortran module
  and a Mathematica interface.
\bigbreak
{\em Typical running time:}
  Dependent on the complexity of the function. GPLs with typlical
  weight up to five evaluate in the millisecond range.
\bigbreak
{\em Limitations:}
  There are no theoretical limitations of the weight through the
  algorithm. However, for arbitrary parameters there are limits
  through runtime for increasing weight.

{\raggedright

}
\end{small}

\newpage

\section{Introduction}
It is well known that analytic calculations of higher-order
corrections in quantum field theory give rise to polylogarithms. In
the calculation of master integrals this usually happens when solving
complicated Mellin-Barnes integrals or differential equations.  For
processes involving many scales, these are not just harmonic
polylogarithms~\cite{Remiddi:1999ew} any more.  Instead, generalised
or Goncharov polylogarithms (GPL) are
required~\cite{Goncharov:1998kja}.

Much effort has been dedicated to harmonic polylogarithms~\cite{
Buehler:2011ev, Gehrmann:2001pz, Huber:2005yg, Maitre:2005uu,
Ablinger:2018sat}, making their numerical evaluation fast and
effortless. Also the more general two-dimensional harmonic
polylogarithms (for a definition see Section~\ref{sec:validation}) can
efficiently be evaluated~\cite{Gehrmann:2001jv}.  Unfortunately, the
same cannot quite be said for generalised polylogarithms. Worse yet,
as we enter the era of high-precision fully-differential NNLO and
N$^3$LO calculations, being able to merely evaluate these functions is
not sufficient anymore. We need to be able to integrate over GPLs
numerically within a Monte Carlo code, meaning that speed ceases to be
just a luxury -- it becomes critical.

There are two public methods that deal with the numeric aspect of
GPLs: a general implementation in the computer algebra system
GiNaC~\cite{Vollinga:2004sn} and a set of reduction rules to reduce
GPLs to known functions~\cite{Frellesvig:2016ske}. The latter is
implemented in Mathematica and can be difficult to use if the choice
of branch cuts matters. The former, written in C++, on the other hand,
can be cumbersome to interface with Monte Carlo programs which are
usually written in Fortran. The computer algebra library GiNaC
performs the numerical evaluation symbolically, resulting in
performance unsuitable for Monte Carlo integration. The algorithm
employed by GiNaC is also implemented in
Maple~\cite{Frellesvig:2018lmm}.

Hence, we present \name{}, an easy-to-use Fortran implementation of
the algorithm presented in~\cite{Vollinga:2004sn}, enjoying the raw
speed of the compiled language's complex number arithmetic without
sacrificing simplicity.

This paper is structured as follows: in Section~\ref{sec:defs} we
formally introduce GPLs and the different notations we are using as
well as some general properties. Next, in Section~\ref{sec:use} we
discuss how to obtain, install and use \name{}. For the inclined
reader, Section~\ref{sec:alg} discusses the algorithm used by GiNaC
and \name{} in detail, providing examples. Finally, we compare the
code's performance on a set of test cases in
Section~\ref{sec:validation} before we conclude in
Section~\ref{sec:conclusion}.


\section{Notation and properties of GPLs}\label{sec:defs}

GPLs are complex-valued functions that depend on $m$ complex
parameters $z_1,...,z_m$ as well as an argument $y$. We can define a
GPL as a nested integral with $z_m\neq 0$
\begin{align}
 G(z_1,...,z_m\ ;y) \equiv
   \int_0^y         \frac{\D t_1}{t_1-z_1}
   \int_0^{t_1    } \frac{\D t_2}{t_2-z_2}
   \cdots
   \int_0^{t_{m-1}} \frac{\D t_m}{t_m-z_m}\,.
 \label{eq:gpl}
\end{align}
Alternatively, they can also be defined in recursive form as
\begin{align}
  G(z_1,...,z_m\ ;y)=\int_0^y \frac{\D t_1}{t_1-z_1}
  G(z_2,...,z_m\ ;t_1)\,,
\end{align}
where the base case of $m=1$ is just a logarithm
\begin{align}
G(z\ ;y)= \log\Big(1-\frac{y}{z}\Big)\,.
\end{align}
To also cover the case of $z_m=0$ we define
\begin{align}
  G(\underbrace{0,...,0}_{m}\ ;y)\equiv G(0_m\ ; y)
     =\frac{(\log y)^m}{m!}\,,
\end{align}
where we denote a string of $m$ zeros as $0_m$.

We call $G(z_1,...,z_m;y)$ \emph{flat} since all parameters are
explicit. However, this notation can be cumbersome if many of the
$z_i$ are zero. In this case we introduce the \emph{condensed}
notation which uses partial weights $m_i$ in order to keep track of
the number of zeros in front of the parameter $z_i$
\begin{align}
  G_{m_1,...,m_k}\big(z_1,...,z_k\ ;y\big) \equiv G\big(
    0_{m_1-1},
    z_1,...,z_{k-1},
    0_{m_k-1},z_k\ ;y\big)\,.
\end{align}
Both notations will be used interchangeably. We say that this GPL is
of depth $k$ as it has $k$ non-zero parameters (not counting $y$). Its
total weight is $m=\sum m_i$.

\subsection{Multiple polylogarithms}

Multiple polylogarithms (MPLs) are a related class of functions that
also generalise logarithms. They are defined as an infinite nested
series
\begin{align}
  \Li_{m_1,...,m_k}(x_1,...,x_k) \equiv
    \sum_{i_1 > \cdots > i_k}^\infty
       \frac{x_1^{i_1}}{i_1^{m_1}} \cdots
       \frac{x_k^{i_k}}{i_k^{m_k}}\,,
  \label{eq:mpl}
\end{align}
where $m_1,...,m_k$ are integer weights. If there is only one argument
present, they reduce to classical polylogarithms $\Li_m(x)$.

MPLs are closely related to GPLs through
\begin{align}
  \Li_{m_1,...,m_k}(x_1,...,x_k) =
    (-1)^k G_{m_1,...,m_k} \Big(
        \frac1{x_1} , \frac1{x_1 x_2} ,...,
        \frac1{x_1 \cdots x_k}\ ;1 \Big)\,.
\end{align}
This can be inverted by performing an iterated substitution
\begin{align}
  u_1 = \frac1{x_1}\,,\quad
  u_2 = \frac1{x_1 x_2} = \frac{u_1}{x_1}\,,
  \quad...\qquad
  u_k = \frac1{x_1 ... x_k} = \frac{u_{k-1}}{x_k}\,,
\end{align}
allowing us to write the GPLs in terms of MPLs
\begin{align}
  G_{m_1,...,m_k}(u_1,...,u_k\ ;1)
      = (-1)^k \Li_{m_1,...,m_k} \Big(
            \frac1{u_1},\frac{u_1}{u_2} , ... ,
                \frac{u_{k-1}}{u_k} \Big)\,.
  \label{eq:rel_gpl_mpl}
\end{align}
In \eqref{eq:rel_gpl_mpl}, the left-hand side is an integral
representation whereas the right-hand side is a series representation.

GPLs with arbitrary parameters satisfy the scaling relation
\begin{align}
  G(z_1,...,z_m\ ;y) = G(\kappa z_1,...,\kappa z_m\ ;\kappa y) \label{eq:scaling}
\end{align}
for any complex number $\kappa \ne 0$.  \eqref{eq:rel_gpl_mpl} assumes
the argument of $G$ is equal to one.  Using the scaling relation we
can normalise $G(z_1,...,z_m;y)$ with $\kappa=1/y$ to guarantee that
the argument is indeed one.

For the numerical evaluation the main idea will be to compute
$G$-functions by reducing them to their corresponding series
representation \eqref{eq:rel_gpl_mpl}.

\subsection{Convergence properties}
If we want to use an infinite series for numerical evaluation of GPLs,
the series needs to be convergent. It can be shown~\cite{
Vollinga:2004sn} that an MPL $\Li_{m_1,...,m_k}(x_1,...,x_k)$ is
convergent if the conditions
\begin{align}
   |x_1 \cdots x_k| < 1 \qquad \text{and} \qquad (m_1,x_1) \ne (1,1)
\end{align}
are satisfied. Using the relation \eqref{eq:rel_gpl_mpl}, this
translates to a sufficient convergence criterion for the integral
representation. We find that if
\begin{align}
  |y| < |z_i|\quad \forall i=1,...,k\quad \text{and} \quad (m_1,y/z_1) \ne (1,1)\,,
  \label{eq:convergence_criterion}
\end{align}
$G_{m_1,...,m_k}(z_1,...,z_k\ ;y)$ is convergent.

In Section~\ref{sec:alg} we will review the algorithm developed
by~\cite{Vollinga:2004sn} to transform any GPL into this form.

\subsection{Shuffle algebra and trailing zeros}
If the last parameter $z_k$ of a GPL $G_{m_1,...,m_k}(z_1,...,z_k\
;y)$ vanishes, the convergence
criterion~\eqref{eq:convergence_criterion} is not fulfilled. Hence,
any algorithm that intents to exploit~\eqref{eq:mpl} for numerical
evaluation needs to remove trailing zeros.

We can exploit the fact that GPLs satisfy two Hopf algebras: a shuffle
algebra and a stuffle
algebra~\cite{Vollinga:2004sn,Frellesvig:2016ske,Duhr:2014woa}. Here,
we will only be needing the former. It allows us to write the product
of two GPLs with parameters $\vec a$ and $\vec b$ as
\begin{align}
  G(\vec a\ ;y) \cdot G(\vec b\ ;y)
    = \sum_{\vec c = \vec a\,\shuffle\,\vec b}
        G(\vec c\ ;y)\,.
  \label{eq:shuffle_algebra}
\end{align}
The sum in the right-hand side of \eqref{eq:shuffle_algebra} runs over
all elements of the shuffle product of the list $\vec a$ with
$\vec b$.  This shuffle product gives the set of all permutations of
the elements in $\vec a$ and $\vec b$ that preserve the respective
orderings of $\vec a$ and $\vec b$. For practical implementations, a
recursive algorithm exists~\cite{Duhr:2019tlz}.

\section{Installation and usage}\label{sec:use}
The code is available in a public GitLab repository hosted by the Paul
Scherrer Institut at
\begin{lstlisting}[language=bash]
    (*@\url{\basegit}@*)
\end{lstlisting}
From this URL a release version can be downloaded in compressed form.
Alternatively, \name{} can be obtained by cloning using the {\tt git}
command
\begin{lstlisting}[language=bash]
    git clone (*@\url{\basegit.git}@*)
\end{lstlisting}
This will download \name{} into a subfolder called {\tt handyG}.
Within this folder
\begin{lstlisting}[language=bash]
    git pull
\end{lstlisting}
can be used to update \name{}.

\subsection{Installation}
\name{} should run on a variety of systems though this can obviously
not be guaranteed. The code follows the conventional installation
scheme\footnote{Despite the name,
\lstinline[language=bash]{./configure} has nothing to do with
autotools.}
\begin{lstlisting}[language=bash]
./configure  # Look for compilers and make a guess at
             # necessary flags
make all     # Compiles the library
make check   # Performs a variety of checks (optional)
make install # Installs library into prefix (optional)
\end{lstlisting}
\name{} has a Mathematica interface (activate with
\lstinline{--with-mcc}) and a GiNaC interface (activate with
\lstinline{--with-ginac}) that can be activated by supplying the
necessary flags to \lstinline{./configure}. The latter is only used
for testing purposes and is not actually required for running. Another
important flag is \lstinline{--quad} which enables quadruple precision
in Fortran. Note that this will slow down \name{}, so that it should
only be used if double-precision is indeed not enough.

The compilation process creates the following results
\\
\begin{tabular}{ll}
{\tt libhandyg.a} & the \name{} library\\
{\tt handyg.mod}  & the module files for Fortran 90\\
{\tt geval}       & a binary file for quick-and-dirty evaluation\\
{\tt handyG}      & the Mathematica interface
\end{tabular}
\bigbreak
An overview of systems on which the code was successfully tested can
be found in Table~\ref{tab:sys} (see Section~\ref{sec:validation} for
performance).

\begin{figure}
\centering
\def\mr#1{\multirow{2}{*}{#1}}
\begin{tabular}{l|l|l|l}
\bf Operating System     & \bf Processor                   & \bf Compiler   & {\tt math} \\\hline
    Scientific Linux 6.0 &     Xeon E3 Sandy Bridge 3.3GHz & gcc 4.4.4$^*$  & N/A    \\\hline
\mr{Scientific Linux 6.4}& \mr{Xeon E5 Broadwell 2.1GHz}   & gcc 8.2.0      & 11.0.0 \\
                         &                                 & intel 14.0.2   & N/A    \\\hline
\mr{Scientific Linux 7.6}& \mr{Xeon E3 Sandy Bridge 3.3GHz}& gcc 8.2.0      & 11.0.0 \\
                         &                                 & intel 19.0.3   & N/A    \\\hline
    Ubuntu 18.04.2       &     i5 Kaby Lake R 1.7GHz       & gcc 7.4.0      & 11.3.0 \\\hline
    macOS 10.12.6        &     i5 Broadwell 1.6GHz         & gcc 5.1.0      & 11.0.1 \\\hline
\mr{macOS 10.14.5}       & Core M Broadwell 0.9GHz         & gcc 8.3.0      & 11.3.0 \\
                         &     i5 Ivy Bridge 2.5GHz        & gcc 8.3.0      & 11.3.0 \\\hline

\end{tabular}

\renewcommand{\figurename}{Table} \caption{An overview of systems
under which \name{} works as expected. All processors are manufactured
by Intel. {\tt math} indicated the version of Mathematica used. The
$*$ indicates that for this version of gcc no optimisation is
available.}
\label{tab:sys}
\end{figure}

\subsection{Usage in Fortran}
\name{} is written with Fortran in mind. We provide a module {\tt
handyg.mod} containing the following objects
\begin{itemize}
    \item {\tt prec}:
    the working precision as a Fortran {\tt kind}. This is read-only,
    the code needs to be reconfigured for a change to take effect.
    Note that this does not necessarily increase the result's
    precision without also changing the next options.

    \item {\tt set\_options}:

    a subroutine to set runtime parameters of \name{}. {\tt
    set\_options} takes the following arguments
    \begin{itemize}
        \item
        \lstinline{real(kind=prec) :: MPLdel = 1e-15}: difference
        between two successive terms at which the series
        expansion~\eqref{eq:mpl} is truncated.

        \item
        \lstinline{integer LiInf = 1000}: number of terms in the
        expansion of classical polylogarithms.

        \item
        \lstinline{real(kind=prec) :: hCircle = 1.1}: the size of the
        H\"{o}lder circle $\lambda$ (see
        Section~\ref{sec:increase_conv_rate}).
    \end{itemize}
    For an example of how to use {\tt set\_options}, see
    Listing~\ref{lst:set_options}.

    \item {\tt inum}:

    a datatype to handle $\io^+$-prescription (see
    Section~\ref{sec:i0}).

    \item {\tt clearcache}:

    \name{} caches a certain number of classical polylogarithms (see
    Section~\ref{sec:cache}). This resets the cache (in a Monte Carlo
    this should be called at every phase space point).

    \item {\tt G}:

    the main interface for generalised polylogarithms.

\end{itemize}
\begin{figure}
\begin{lstlisting}
real(kind=prec) :: delta, circle
integer inf
delta = 1e-15
inf = 1000
circle = 1.1
call set_options(MPLdel  = delta , &
                 LiInf   = inf   , &
                 hCircle = circle)
\end{lstlisting}
\renewcommand{\figurename}{Listing}
\caption{The default values of the options of \name}
\label{lst:set_options}

\end{figure}
\begin{figure}
\begin{lstlisting}
PROGRAM gtest
  use handyG
  complex(kind=prec) :: res(5), x, weights(4)
  call clearcache

  x = 0.3 ! the parameter

  ! flat form with integers
  res(1) = G((/ 1, 2, 1 /))

  ! very flat form for real numbers using F2003 arrays
  res(2) = G([ 1., 0., 0.5, real(x)])
  ! this is equivalent to the flat expression
  res(2) = G([ 1., 0., 0.5 ], real(x))
  ! or in condesed form
  res(2) = G((/1, 2/), (/ 1., 0.5 /), real(x))

  ! flat form with complex arguments
  weights = [(1.,0.), (0.,0.), (0.5,0.), (1.,1.) ]
  res(3) = G(weights, x)

  ! flat form with explicit i0-prescription
  res(4) = G([inum(1.,+1),inum(0,+1),inum(5,+1)], &
                    inum(1/x,di0))
  res(5) = G([inum(1.,-1),inum(0,+1),inum(5,+1)],&
                    inum(1/x,di0))
  ! this is equivalent to
  res(5) = G((/1,2/),[inum(1.,-1),inum(5,+1)], &
                    inum(1/x,+1))

  do i =1,5
    write(*,900) i, real(res(i)), aimag(res(i))
  enddo
900 FORMAT("res(",I1,") = ",F9.6,"+",F9.6,"i")  
END PROGRAM gtest
\end{lstlisting}
\renewcommand{\figurename}{Listing}
\caption{The example program {\tt example.f90} to calculate the
example in \eqref{eq:example}}
\label{lst:example.f90}
\end{figure}

In Listing~\ref{lst:example.f90} we show an example program to
calculate the following GPLs
\begin{align}\begin{split}
{\tt res}(1) &= G(1,2;1)\,,\\
{\tt res}(2) &= G\big(1,0,\tfrac12   ;x\big) = G_{1,2  }\big(1,\tfrac12   ;x\big)\,,\\
{\tt res}(3) &= G\big(1,0,\tfrac12,1+\I;x\big) = G_{1,2,1}\big(1,\tfrac12,1+\I;x\big)\,,\\
{\tt res}(4) &= G\big(1_+,0,5;\tfrac1x\big)\,,\\
{\tt res}(5) &= G\big(1_-,0,5;\tfrac1x\big)\,,
\end{split}\label{eq:example}\end{align}
with $x=0.3$ and $1_\pm$ indicating $1\pm\io^+$.

The easiest way to compile the code is with {\tt pkg-config}. Assuming
\name{} has been installed with {\tt make install}, the example
program {\tt example.f90} can be compiled as (assuming you are using
GFortran)
\begin{lstlisting}[language=bash]
(*@\$@*) gfortran -o example example.f90 \
       `pkg-config --cflags --libs handyg`
(*@\$@*) ./example
res(1) = -0.822467+ 0.000000i
res(2) =  0.128388+ 0.000000i
res(3) = -0.003748+ 0.003980i
res(4) = -0.961279+-0.662888i
res(5) = -0.961279+ 0.662888i
\end{lstlisting}
If {\tt pkg-config} is not available and/or for non-standard
installations it might be necessary to specify the search
paths\footnote{Some versions of GFortran specify a search path for
modules. {\tt ifort} does this automatically.}
\begin{lstlisting}[language=bash]
(*@\$@*) gfortran -o example example.f90 \
>     -I/absolute/path/to/handyG -fdefault-real-8 \
>     -L/absolute/path/to/handyG -lhandyg
\end{lstlisting}

\begin{figure}
\begin{lstlisting}[language=Mathematica]
In[1]:= Install["handyG"];
handyG by L. Naterop, Y. Ulrich, A. Signer

In[2]:= x=0.3;

In[3]:= res[1] = G[1,2,1]

Out[3]= -0.822467

In[4]:= res[2] = G[1,0,1/2,x]

Out[4]= 0.128388

In[5]:= res[3] = G[1,0,1/2,1+I,x]

Out[5]= -0.003747969 + 0.00398002 I

In[6]:= res[4] = G[(*@1$_+$@*),5,1/x]

Out[6]= -1.12732 - 0.701026 I

In[7]:= res[5] = G[(*@1$_-$@*),5,1/x]

Out[7]= -1.12732 + 0.701026 I

\end{lstlisting}
\renewcommand{\figurename}{Listing}
\caption{An example of how to use \name{} in Mathematica to calculate
the functions of \eqref{eq:example}.}
\label{lst:example.m}
\end{figure}

\clearpage

\subsection{Usage in Mathematica}
Mathematica is arguably one of the most used computer algebra system
among particle physicists. Hence, we have interfaced our code to
Mathematica using Wolfram's MathLink interface (for a review on how
this works, see~\cite{Hahn:2011gf}). In Listing~\ref{lst:example.m} we
show how to calculate the functions in~\eqref{eq:example} in
Mathematica, assuming that the code was installed with {\tt make
install}. The subscript $1_\pm$, indicating the side of the branch
cut, can be entered using {\tt SubPlus} ({\tt SubMinus}) or using
\mmaCU and \mmaPlus, (\mmaCU and \mmaMinus).  When using \name{} in
Mathematica, keep in mind that it uses Fortran which means that
computations are performed with fixed precision.


\subsection{Proper \texorpdfstring{$\io^+$}{i0+-} prescription}
\label{sec:i0}
To evaluate integrals in the physical kinematic region, we often need
to prescribe on which side of any potential branch cut a parameter
lies. This is done by adding an infinitesimal imaginary part to the
parameter. In \name{} this is implemented using a custom data
type\footnote{Note that, due to padding, the actual size of {\tt inum}
may be as large as 24 byte.} that keeps track of both the (potentially
complex) number {\tt c} and the sign of the imaginary part {\tt i0}
\begin{lstlisting}
  type inum
    complex(kind=prec) :: c
    integer(1) :: i0
  end type inum
\end{lstlisting}
There are a few constants and procedures implemented for the user's
convenience
\begin{lstlisting}
  integer(1), parameter :: di0 = +1
  type(inum), parameter :: izero=inum( 0.,di0)

  FUNCTION TOINUM(z, s)
  real(kind=prec) :: z(:)
  type(inum) :: toinum(size(z))
  integer(1),optional :: s
  ...
  END FUNCTION TOINUM

  FUNCTION TOCMPLX
  type(inum) :: z
  complex(kind=prec) tocmplx
  ...
  END FUNCTION TOCMPLX
\end{lstlisting}
The variable \lstinline{di0} specifies the default imaginary part that
will be used if nothing is specified explicitly. The functions
\lstinline{toinum} and \lstinline{tocmplx} can be used to convert
lists and numbers to \lstinline{inum} objects and complex numbers,
respectively.

Finally, \lstinline{real}, \lstinline{aimag} and \lstinline{abs} work
as expected even on objects of type \lstinline{inum}.


\subsection{Cache system}\label{sec:cache}
\name{} has a cache system for classical polylogarithms. This is
controlled through the parameter
\begin{lstlisting}
integer, parameter :: PolyLogCacheSize(2) = (/ (*@$n$@*), (*@$m_\text{max}$@*) /)
\end{lstlisting}
in {\tt globals.f90}. This caches $n$ polylogarithms of the form
$\Li_m(x)$ for $2\le m\le m_\text{max}$ each. The default values are
$n=100$ and $n_\text{max}=5$.

The cache system consumes
\begin{align*}
 n\times m_\text{max}\times \big( 2\times
        \text{\lstinline{sizeof(complex(kind=prec))}}
     + 1\text{byte} + \text{padding} \big)
 = 12\,\mathrm{kB}
\end{align*}
bytes of memory in the default settings. This is a very small price to
pay for improving the evaluation speed considerably.

The gain from a similar system for convergent MPLs or even entire GPLs
is presently not worth the effort.


\section{The algorithm}\label{sec:alg}

The central idea to numerically evaluate GPLs is to first map their
parameters to the domain where the corresponding series representation
is convergent \eqref{eq:convergence_criterion} and to then use the
series expansion up to some finite order. Thus, we will first look at
how to remove trailing zeros in
Section~\ref{sec:remove_trailing_zeros}, and then how to make a GPL
without trailing zeros convergent in Section~\ref{sec:make_convergent}
as presented in~\cite{Vollinga:2004sn}. In
Section~\ref{sec:increase_conv_rate}, we comment on accelerating the
convergence of already convergent GPLs. Finally, in
Section~\ref{sec:example_reduction} we apply the algorithm to an
explicit example.


\subsection{Removal of trailing zeros}
\label{sec:remove_trailing_zeros}

Consider a GPL of weight $m$ with $m\!-\!j$ trailing zeros
\begin{align*}
G(z_1,...,z_j,0_{m-j}\ ;y)\,.
\end{align*}
We now shuffle $\vec a = (z_1,...,z_j,0_{m-j-1})$ with $\vec b = (0)$.
This results in $m\!-\!j$ times the original GPL as well as terms with
less trailing zeros
\begin{align}\begin{split}
  G(0\ ;y)\cdot G(z_1,...,z_j,0_{m-j-1}\ ;y)
     &= (m-j) G(z_1,...,z_j,0_{m-j}\ ;y)  \\&\qquad
        + \sum_{\vec s} G(s_1,...,s_j,z_j,0_{m-j-1}\ ;y)\,,
\end{split}\label{eq:trail}\end{align}
where the sum runs over all shuffle $\vec s=(z_1,...,z_{j-1})\,
\shuffle\, (0)$. We now solve~\eqref{eq:trail} for
$G(z_1,...,z_j,0_{m-j};y)$ and obtain an expression with fewer
trailing zeros. By applying this strategy recursively, we can remove
all trailing zeros.

\subsection{Making GPLs convergent}
\label{sec:make_convergent}

\subsubsection{Reduction to pending integrals}
Consider a GPL of the form
\begin{align}
  G(a_1,...,a_{i-1},s_r,a_{i+1},...,a_m\ ;y)
  \label{eq:non_conv_G}
\end{align}
where $s_r(=a_i)$ has the smallest absolute value among all the
non-zero parameters in $G$. If $|s_r| < |y|$, \eqref{eq:non_conv_G}
has no convergent series expansion.  In order to remove the smallest
weight $s_r$, we apply the fundamental theorem of calculus to generate
terms where $s_r$ is either integrated over or not present anymore
\begin{align}
\begin{split}
  G(a_1,...,a_{i-1},s_r,a_{i+1},...,a_m\ ;y) =
  G(a_1,...,a_{i-1}, 0 ,a_{i+1},...,a_m\ ;y)   \\
 +  \int_0^{s_r} \D s_{r+1} \frac{\partial}{\partial s_{r+1}}
  G(a_1,...,a_{i-1},s_{r+1},a_{i+1},...,a_m\ ;y)\,.
\end{split}
\end{align}
For the second term we use partial fraction decomposition and
integration by parts. Then we obtain different results depending on
where $s_r$ is in the parameter list:

\begin{itemize}

\item
\emph{If $s_r$ appears first} in the list (i.e. $i = 1$ and
$s_r=a_1$) we find
\begin{align}
\begin{split}
  G(s_r,a_{i+1},...,a_m\ ;y) =
  G(0  ,a_{i+1},...,a_m\ ;y) +
    \underbrace{\int_0^{s_r}\frac{\D s_{r+1}}{s_{r+1}-y     }}_{G(y\ ;s_r)} G(a_{i+1},...,a_m\ ;y) \\
  + \underbrace{\int_0^{s_r}\frac{\D s_{r+1}}{s_{r+1}-a_{i+1}} G(s_{r+1},a_{i+2},..,a_m\ ;y)}_{\text{pending integral}}
  - \underbrace{\int_0^{s_r}\frac{\D s_{r+1}}{s_{r+1}-a_{i+1}}}_{G(a_2\ ;s_r)} G(a_{i+1},...,a_m\ ;y)
\,.
\end{split}
\label{eq:s_r_at_beginning}
\end{align}
In the first term on the right-hand side, $s_r$ is absent. Therefore
the resulting GPL is simpler. It might still be non-convergent, but we
can use this method recursively on the resulting GPLs until we end up
with convergent GPLs.

In the second and fourth terms the integration variable $s_{r+1}$ does
not appear in the parameters of the GPL, so that the integral can be
solved (we write the solution as a GPL instead of a logarithm to be
able to continue recursively).

The third term does have the integration variable $s_{r+1}$ among the
weights and therefore yields what we refer to as a pending integral.
This object can be written as a linear combination of simpler GPLs as
we will see in Section~\ref{sec:eval_pend_int}.

Note that all GPLs on the right-hand side have depth reduced by one.

\item

\emph{If $s_r$ appears in the middle} of the list, i.e. $1 < i < m$,
we find
\begin{align}
\begin{split}
    G(a_1,...,a_{i-1},s_r, a_{i+1},...,a_m\ ;y) =  & \\
  + G(a_1,...,a_{i-1}, 0 ,&a_{i+1},...,a_m\ ;y)      \\
  -             \int_0^{s_r}  \frac{\D s_{r+1}}{s_{r+1}-a_{i-1}}                     & G(a_1,...,a_{i-2},s_{r+1},a_{i+1},...,a_m\ ;y) \\
  + \underbrace{\int_0^{s_r}  \frac{\D s_{r+1}}{s_{r+1}-a_{i-1}}}_{G(a_{i-1}\ ;s_r)} & G(a_1,...,a_{i-1},        a_{i+1},...,a_m\ ;y) \\
  +             \int_0^{s_r}  \frac{\D s_{r+1}}{s_{r+1}-a_{i+1}}                     & G(a_1,...,a_{i-1},s_{r+1},a_{i+2},...,a_m\ ;y) \\
  - \underbrace{\int_0^{s_r}  \frac{\D s_{r+1}}{s_{r+1}-a_{i+1}}}_{G(a_{i+1}\ ;s_r)} & G(a_1,...,a_{i-1},        a_{i+1},...,a_m\ ;y)
\,.
\end{split}
\label{eq:s_r_in_middle}
\end{align}
Again we obtain simpler GPLs (without $s_r$ or lower depth) as well as
pending integrals.

\item
\emph{If $s_r$ appears last} in the list, i.e. $i = m$, we use the
shuffle algebra to remove $s_r$ from the last place, just as we have
done to remove trailing zeros.

\end{itemize}

We repeat these steps also for GPLs that are already under a pending
integral.

\subsection{Evaluation of pending integrals}
\label{sec:eval_pend_int}

The most general term created by the procedure of the last section is
of the form
\begin{align}\begin{split}
  \PI\Big(\vec p=(y',\vec b),i,\vec g = (\vec a,y)\Big)
  &\equiv
      \int_0^{y'     } \frac{\D s_1}{s_1-b_1}
      \int_0^{s_1    } \frac{\D s_2}{s_2-b_2} \cdots
      \int_0^{s_{r-1}} \frac{\D s_r}{s_r-b_r}
      \\&\qquad
            G(a_1,...,a_{i-1},s_r,a_{i+1},...,a_m\ ;y) \, .
\end{split}
\end{align}
Here we have adopted the convention that $i=0$ implies that the
integration variable does not appear inside the GPL. For example
\begin{align*}\begin{split}
\PI\Big(\vec p=(1,2,3),0,(4,5)\Big)&= \int_0^1 \frac{\D s_1}{s_1-2}\int_0^{s_1}\frac{\D s_2}{s_2-3} G(4;5) \, \\
\PI\Big(\vec p=(1,2,3),2,(4,5)\Big)&= \int_0^1 \frac{\D s_1}{s_1-2}\int_0^{s_1}\frac{\D s_2}{s_2-3} G(4,s_2;5)\,.
\end{split}\end{align*}

As we use the algorithm, we need a way to collapse the pending
integrals back down again. As an example, consider the case $i=1$
\begin{align}
\begin{split}
  & \PI\Big(\vec p=(y',\vec b),1,\vec g=(\vec a,y)\Big)
    =              \int_0^{y'} \frac{\D s_1}{s_1-b_1}\cdots\int_0^{s_{r-1}} \frac{\D s_r}{s_r-b_r}                     G(s_r,a_{i+1},...,a_m\ ;y) = \\
  &    \underbrace{\int_0^{y'} \frac{\D s_1}{s_1-b_1}\cdots\int_0^{s_{r-1}} \frac{\D s_r}{s_r-b_r}}_{\PI(\vec p,0,())} G( 0, a_{i+1},...,a_m\ ;y) \\
  & +  \underbrace{\int_0^{y'} \frac{\D s_1}{s_1-b_1}\cdots\int_0^{s_{r-1}} \frac{\D s_r}{s_r-b_r} \int_0^{s_r}\frac{\D s_{r+1}}{s_{r+1}-y}}_{\PI\Big((\vec p,y),0,()\Big)}  G(a_{i+1},...,a_m\ ;y) \\
  & +  \underbrace{\int_0^{y'} \frac{\D s_1}{s_1-b_1}\cdots\int_0^{s_{r-1}} \frac{\D s_r}{s_r-b_r} \int_0^{s_r}\frac{\D s_{r+1}}{s_{r+1}-a_{i+1}}  G(s_{r+1},a_{i+2},...,a_m\ ;y) }_{\PI\Big((\vec p,a_{i+1}),1,(a_{i+2},...,a_m;y)\Big)}\\
  & -  \underbrace{\int_0^{y'} \frac{\D s_1}{s_1-b_1}\cdots\int_0^{s_{r-1}} \frac{\D s_r}{s_r-b_r} \int_0^{s_r}\frac{\D s_{r+1}}{s_{r+1}-a_{i+1}}}_{\PI\Big((\vec p,a_{i+1}),0,()\Big)}  G(a_{i+1},...,a_m\ ;y)
\\&= \PI\Big(\vec p, 0, ()\Big) G(0,a_{i+1},...,a_m\ ; y)
   + \PI\Big((\vec p, y), 0, () \Big) G(a_{i+1}, ...., a_m\ ; y)\\&
   + \PI\Big((\vec p, a_{i+1}), 1, (a_{i+2}, ..., a_m\ ; y)\Big)
   - \PI\Big((\vec p, a_{i+1}), 0, ()\Big) G(a_{i+1}, ..., a_m\ ; y)\,.
\end{split}
\end{align}
The other combinations follow similarly
\begin{align}
\begin{split}
   \PI\Big( \vec p         , i ,(\vec{a};y)                        \Big) &=
 + \PI\Big( \vec p         , 0 ,()                                 \Big) \, G(a_1,...,a_{i-1},0,a_{i+1},...,a_m\ ;y) \\&
 - \PI\Big((\vec p,a_{i-1}),i-1,(a_{i+1},...,a_m; y)               \Big) \\&
 + \PI\Big((\vec p,a_{i-1}), 0 ,()                                 \Big) \, G(a_1,...,a_{i-1},  a_{i+1},...,a_m\ ;y) \\&
 + \PI\Big((\vec p,a_{i+1}), i ,(a_1,...,a_{i-1},a_{i+2},...,a_m;y)\Big) \\&
 - \PI\Big((\vec p,a_{i+1}), 1 ,()                                 \Big) \, G(a_1,...,a_{i-1},  a_{i+1},...,a_m\ ;y)\,.
\end{split}
\end{align}

As we recursively apply the algorithm, we increase the number of
pending integrals in front but decrease the depth of the $G$-functions
by one unit in every recursion step. We do this until
\begin{enumerate}
\item[(a)]
the only GPLs remaining under pending integrals are of depth one, i.e.
$G_m(s_r\; y)$,

\item[(b)]
$s_r$ is the argument, i.e. $G(...\ ;s_r)$, or

\item[(c)]
there are no GPLs under pending integrals.

\end{enumerate}
We now discuss all these cases in turn:

\begin{enumerate}

\item[(a)]
For GPLs of depth one, i.e. $G_m(s_{r \pm}; y)$, we will be working
with explicit logarithms.  Hence, we need to indicate the
infinitesimal imaginary part.  We have to distinguish two cases: $m=1$
and $m>1$. For $m = 1$ we have
\begin{align}
G_1(s_{r \pm};y) = G_1(y_{2 \mp};s_r) - G(0;\ s_r) + \log(-y)\,.
\end{align}
Note that we will most likely have pending integrals in front, thus
each term gives again a simpler pending integral
\begin{align}
  \PI\Big(\vec p=(y_\pm', \vec b), 1, (y)    \Big) =
  G(\vec b, y_\mp\ ; y') - G(\vec b, 0\ ; y')
+ \log(-y_\mp) G\big(\vec b, y')
\end{align}
The first and second terms have been reduced to case~(b) and the third
term to case~(c).

For $m > 1$, we note
\begin{align}
G_m(s_{r \pm}\ ; y) = -\zeta(m)
    + \int_0^ {y}  \frac{\D t}{t}G_{m-1}(t_{\pm}\ ; y)
    - \int_0^{s_r} \frac{\D t}{t}G_{m-1}(t_{\pm}\ ; y)\,.
  \label{eq:depth_one_highter_m}
\end{align}
The second and third terms are now longer pending integrals, albeit
with reduced weight
\begin{align}
\begin{split}
           \PI\Big( \vec{p}   , m ,(0_{m-1},y)\Big) &=
 - \zeta(m)\PI\Big( \vec{p}   , 0 ,(         )\Big) \\&\qquad
 +         \PI\Big(( y,0     ),m-1,(0_{m-2};y)\Big)
           \PI\Big( \vec{p}   , 0 ,(         )\Big) \\&\qquad
 -         \PI\Big((\vec{p},0),m-1,(0_{m-2};y)\Big)\,.
\end{split}
\end{align}

\item[(b)]

In this case we end up simply with one large GPL
\begin{align}
\int_0^{y' }     \frac{\D s_1}{s_1-b_1}  \cdots
\int_0^{s_{r-1}} \frac{\D s_r}{s_r-b_r} \, G(\vec a\ ;s_r) =
G((\vec b,\vec a)\ ; y' )\,.
\end{align}
In terms of pending integrals this is written as
\begin{align}
\PI\Big(\vec p=(y',\vec b), m+1, \vec g\Big) = G(\vec b, \vec g\ ; y')\,.
\end{align}

\item[(c)]

If there is no GPL under the pending integral, the integral evaluates
to a GPL
\begin{align}
  \int_0^{y' }     \frac{\D s_1}{s_1-b_1} \cdots
  \int_0^{s_{r-1}} \frac{\D s_r}{s_r-b_r} = G(b_1,...,b_r\ ;y' )\,.
\end{align}

\end{enumerate}

In each case we end up with GPLs that are simpler in the sense that
$s_r$ has been eliminated. These might still be non-convergent due to
other (non-zero) $z_i$ elements being smaller in absolute value than
$y$. But applying the removal of $s_r$ recursively we can eliminate
all $z_i$ for which $|z_i| < |y|$. Therefore in the end we always
obtain convergent GPLs.

\subsection{Increase rate of convergence}
\label{sec:increase_conv_rate}

Even though we have now only convergent GPLs, that does not imply that
the convergence is fast enough for numerical applications.  From now
on we will only consider $y=1$, as we can normalise any convergent GPL
using~\eqref{eq:scaling}. Convergence of such a GPL is slow if some
$z_i$ is close to the unit circle, i.e.
\begin{align}
1 \le |z_i| \le \lambda < 2\,,
\end{align}
where $\lambda$ is a parameter to be chosen.

Only for such $z_i$ we apply the following strategy:
to increase the rate of convergence we can use the fact that GPLs
satisfy the H\"{o}lder convolution equation~\cite{Borwein:1999js}
\begin{align}
   G(z_1,...,z_k\ ;1) = \sum_{j=0}^k (-1)^j
      G\Big(1-z_j,...,1-z_1\ ;1-\tfrac1p\Big)
      G\Big(z_{j+1},...,z_k\ ;  \tfrac1p\Big)\,,
\end{align}
where $p$ is an arbitrary non-zero complex number. Separating the
first and the last term of this sum we obtain for $p=2$ and again
normalising the GPLs on the right-hand side
\begin{align}
G(z_1,...,z_k\ ;1)& =
 G\big(2   z_1 ,...,2   z_k \ ; 1\big) + (-1)^k
 G\big(2(1-z_k),...,2(1-z_1)\ ; 1\big) \\&
 +\sum_{j=1}^{k-1}(-1)^j G\Big(2(1-z_j),...,2(1-z_1)\ ; 1\Big)
                         G\Big(2 z_{j+1},...,2z_k   \ ; 1\Big)
\,.
\end{align}
The first term has now better convergence as all parameters are twice
as big. The GPL appearing in the sum all have reduced weight and are
therefore not relevant for the present discussion.

The second term may or may not be convergent. If not, we repeat the
algorithm outlined in Section~\ref{sec:make_convergent}, including if
necessary, H\"{o}lder convolution. At this stage it is not obvious why
this recipe does indeed lead to a final answer and not to an infinite
recursion. This can be shown by noting that the algorithm does only
replace parameters with zero or permutes them; it does not introduce
new non-trivial parameters. By carefully considering all possible
behaviours under transformation $z\mapsto 2(1-z)$,
\cite{Vollinga:2004sn} proved that this method indeed works.

The choice of $\lambda$ is a trade-off between accuracy and speed. A
typical choice would be $\lambda=1.1$ which is the default in \name{}.
$\lambda$ can be changed using the \lstinline{hCircle} option in
\lstinline{set_options}.

\subsection{An example reduction}
\label{sec:example_reduction}

To illustrate the various aspects discussed so far, we include here an
example of how the algorithm works in practice. For this purpose we
reduce $G(1,0,3;2)$ according to this algorithm until we end up with
logarithms, polylogarithms and convergent MPLs. In our notation of a
non-convergent
GPL we have
\begin{align}
  G(\underbrace{1}_{s_r},
    \underbrace{0}_{a_2},
    \underbrace{3}_{a_3}\ ;
    \underbrace{2}_{y})
  = G(0,0,3;2) + \int_0^1 \D s_1 \frac{\partial}{\partial s_1}G(s_1,0,3;2)
\,.  \label{eq:algo_1}
\end{align}
The first term corresponds to $G_3(3;2)$ and therefore it is a
convergent trilogarithm. The second term has $s_r$ appearing at the
first place. Using~\eqref{eq:s_r_at_beginning} we obtain for the
second term
\begin{align}\begin{split}
  \int_0^1 \D s_1 \frac{\partial}{\partial s_1}G(s_1,0,3\ ;2)
 &= \int_0^1 \frac{\D s_1}{s_1-2}G(0,3\ ;2)
  + \int_0^1 \frac{\D s_1}{s_1-0}G(s_1,3\ ;2)\\&
  - \int_0^1 \frac{\D s_1}{s_1-0} G(0,3\ ;2)\,.
\end{split}  \label{eq:algo_2}
\end{align}
The first and last terms are both conventional functions. Hence, we
only need to worry about the second term which involves a pending
integral. In order to evaluate it, we apply
again~\eqref{eq:s_r_at_beginning} to the GPL under the pending
integral to find
\begin{align}\begin{split}
  G(s_1,3\ ;2) &=
                                 G(0,  3\ ;2)
 +\int_0^{s_1} \frac{\D s_2}{s_2-2}G(  3\ ;2)
 +\int_0^{s_1} \frac{\D s_2}{s_2-3}G(s_2\ ;2) \\&
 -\int_0^{s_1} \frac{\D s_2}{s_2-3}G(  3\ ;2)
\, .
 \end{split} \label{eq:algo_3}
\end{align}
Substituting this back into \eqref{eq:algo_2} gives
\begin{align}
\begin{split}
 \int_0^1 \frac{\D s_1}{s_1-0}&G(s_1,3\ ;2)
=\int_0^1 \frac{\D s_1}{s_1  }G(  0,3\ ;2)
+\int_0^1 \frac{\D s_1}{s_1  }\int_0^{s_1} \frac{\D s_2}{s_2-2}G( 3 \ ;2)\\&
+\int_0^1 \frac{\D s_1}{s_1  }\int_0^{s_1} \frac{\D s_2}{s_2-3}G(s_2\ ;2)
-\int_0^1 \frac{\D s_1}{s_1  }\int_0^{s_1} \frac{\D s_2}{s_2-3}G( 3 \ ;2 )
\,.
\end{split}
\label{eq:algo_4}
\end{align}
Here only the third term is interesting, as the others are
(poly)logarithms. The third term is a pending integral over a GPL of
depth one. Thus,
\begin{align}
\begin{split}
\int_0^1 \frac{\D s_1}{s_1} \int_0^{s_1}&\frac{\D s_2}{s_2-3} G(s_2\ ;2) \\&=
\int_0^1 \frac{\D s_1}{s_1} \int_0^{s_1} \frac{\D s_2}{s_2-3}
  \Big(G(2\ ;s_2) - G(0\ ;s_2) +  \log(-2) \Big)
\end{split}
  \label{eq:algo_5}
\end{align}
The first two terms have $s_r$ as the argument and hence they are
GPLs. The last term is independent of $s_r$, making the integration
trivial.  Unfortunately, the second term $G(0,3,0;1)$ has a trailing
zero. To remove it, we shuffle $G(0,3;1)$ with $G(0;1)$ to find
\begin{align}
G(0,3\ ;1)G(0\ ;1)
    = \sum_{\vec c = (0,3) \shuffle (0)} G(\vec c\ ;1)
    = G(0,3,0\ ;1) + 2\times G(0,0,3\ ;1)\,,
  \label{eq:algo_7}
\end{align}
which we solve for $G(0,3,0\ ;1)$.

Gathering all terms we obtain with $G(0;\ 1)=\log1=0$
\begin{align}
\begin{split}
G(1,0,3\ ;2) &=
    \underbrace{G(0,0,3\ ;2)}_{-\Li_3(2/3)}
  + \underbrace{G(  2  \ ;1)}_{\log(1/2)} \underbrace{G(0,3\ ;2)}_{-\Li_2(2/3)}
  - \cancel{    G(  0  \ ;1) G(0,3\ ;2)}\\&
  + \cancel{    G(  0  \ ;1) G(0,3\ ;2)}
  + \underbrace{G(  0,2\ ;1)}_{-\Li_2(1/3)} \underbrace{G(3\ ;2)}_{\log(1/3)}
  - \underbrace{G(  0,3\ ;1)}_{-\Li_2(2/3)} \underbrace{G(3\ ;2)}_{\log(1/3)} \\&
  + \underbrace{G(0,3,2\ ;1)}_{ \Li_{2,1}(1/3,3/2)}
  + \underbrace{G(  0,3\ ;1)}_{-\Li_2(1/3)} \log(-2)
  - \cancel{    G(  0  \ ;1) G(  0,3\ ;1)} \\&
  + 2 \underbrace{G(0,0,3;1)}_{-\Li_3(1/3)} = -0.81809 - 1.15049\I
\,.
\end{split}
\end{align}


\section{Validation and performance}\label{sec:validation}
The purpose of the this code is to provide a tool for the fast numerical
evaluation of generic GPLs. This is achieved through an `on-the-fly'
reduction of GPLs. For certain subclasses such as (harmonic)
polylogarithms there are obviously faster tailored
routines~\cite{Gehrmann:2001pz,Ablinger:2018sat}.  A particularly
important subclass are \emph{two-dimensional harmonic polylogarithms},
i.e. GPLs where all $z_i\in\{0,+1,1-x,-x\}$.  Up to weight $m=4$,
these objects can be evaluated using the public code {\tt
tdhpl}~\cite{Gehrmann:2001jv}.  {\tt tdhpl} uses hard-coded reduction
rules.

We have validated \name{} for some practical examples of GPLs, namely

\begin{enumerate}\setcounter{enumi}{-1}
    \item
    the GPLs entering the heavy-to-light form factor with full mass
    dependence after simplification~\cite{Engel:2018fsb} (540 GPLs up
    to weight four),

    \item
    all GPLs appearing in the master integrals
    computed~\cite{Chen:2018dpt} for the heavy-to-light form factor
    (1399 functions up to weight four, including the 540 above),

    \item
    the planar integrals for muon-electron scattering in the
    unphysical region $s<0$ and $t<0$ with vanishing electron
    mass~\cite{Mastrolia:2017pfy} (198 functions up to weight four),

    \item
    the non-planar integrals for the same
    process~\cite{DiVita:2018nnh} (1732 GPLs up to weight four),

    \item
    the integrals for Bhabha scattering~\cite{Czakon:2004wm} expressed
    entirely in terms of GPLs~\cite{Henn:2013woa}, and

    \item
    several ten million `random' $G(z_1,...,z_k\,;\,1)$, mimicking physical
    situations. We generate a random list of 110 possible weights (10
    zeros, 50 real and 50 complex entries, all with $|z_i|<3$). Using
    this list we randomly select weights for the GPLs up to $m=5$.
\end{enumerate}
In all four test cases we find complete agreement with
GiNaC\footnote{In some rare cases, depending on the GiNaC installation
{\tt make check} may still fail.}.

\begin{figure}
\centering
\begin{tikzpicture}
    \node [anchor = south]at (0,0){\includegraphics[width=0.8\textwidth]{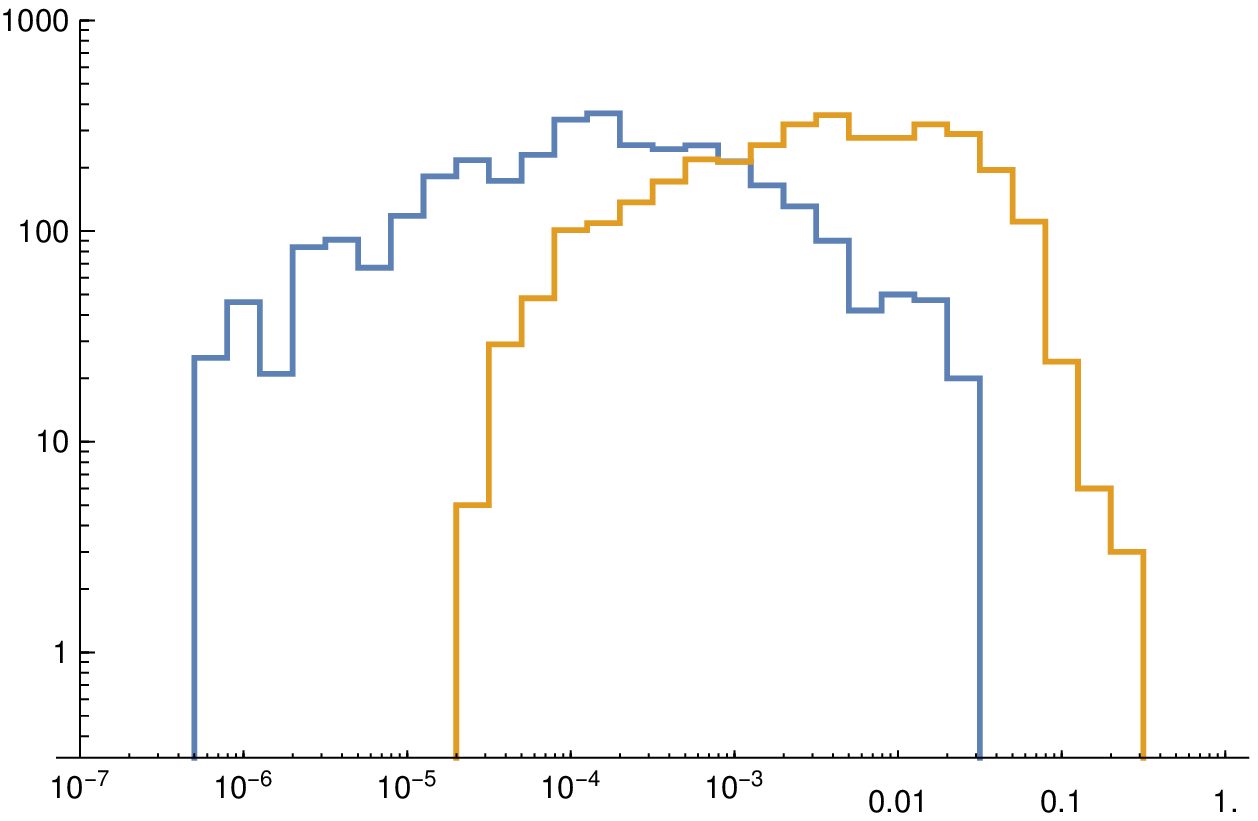}};
    \node at (0,0){$t_{ev}/s$};
    \node [rotate=90]at (-5.5,4){No. of GPLs};
\end{tikzpicture}

\caption{Histogram of average evaluation time of the GPLs needed
in~\cite{Chen:2018dpt,Mastrolia:2017pfy,DiVita:2018nnh} using \name{}
(blue) and GiNaC (yellow)}
\label{fig:chist}
\end{figure}

Of course the speed of any numerical routine strongly depends on the
complexity of the requested function. Hence, comparing total runtime,
while important, does not provide many insights. Instead, one should
study how \name{} and GiNaC perform for different GPLs. This is done
in Figure~\ref{fig:chist}, where we have calculated a total of 3329
GPLs, using both GiNaC and \name{} and histogrammed the average
evaluation time of five successive calls. On average our code is
approximately twenty times faster ($1100\,\text{GPL}/{\rm s}$ v.
$60\,\text{GPL}/{\rm s}$). However, one should keep in mind that the
GiNaC implementation was never intended to be directly used in a Monte
Carlo~\cite{ginac}. Instead, GiNaC would generate C code that
evaluates expressions using double precision. Of course this is only
possible for elementary functions that are implemented in C and not
for, say, GPLs. \name{} fills this gap by providing a low-level
implementation of GPLs suitable for Monte Carlo applications.

\begin{figure}
\centering
\begin{tikzpicture}
    \node [anchor = south]at (0,0){\includegraphics[width=0.8\textwidth]{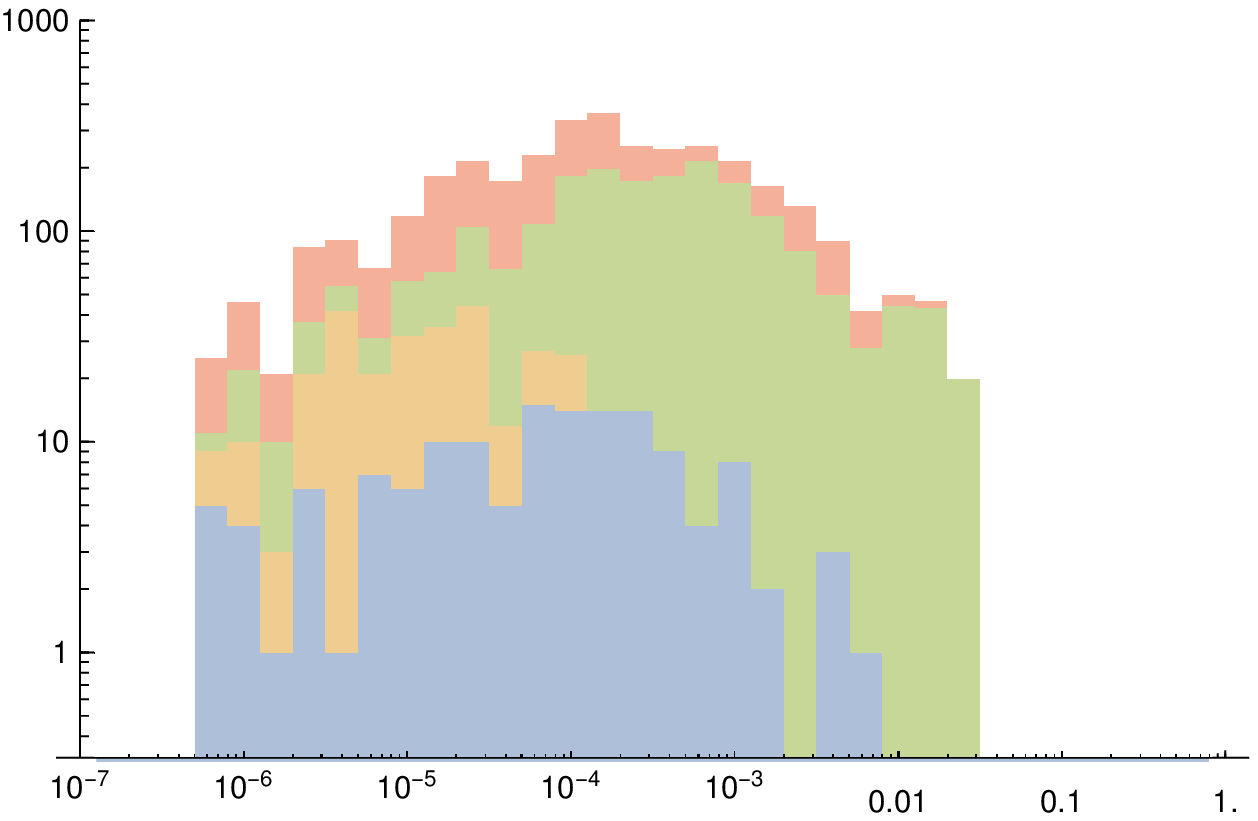}};
    \node at (0,0){$t_{ev}/s$};
    \node [rotate=90]at (-5.5,4){No. of GPLs};
\end{tikzpicture}

\caption{Histogram of average evaluation time of the GPLs using
\name{} broken down by their source: Bhabha scattering (4, blue),
planar $\mu\!-\!e$ scattering (2, yellow), non-planar $\mu\!-\!e$
scattering (3, green) and the heavy-to-light form factor (1, red) }
\label{fig:shist}
\end{figure}

Additionally, we studied in Figure~\ref{fig:shist} how the different
sets of GPLs in the list above compare. For the muon-electron
scattering case, it is perhaps unsurprising that the planar integrals
give rise to easier GPLs than the non-planar integrals.

\begin{figure}
\centering
\begin{tikzpicture}
    \node [anchor = south]at (0,0){\includegraphics[width=0.8\textwidth]{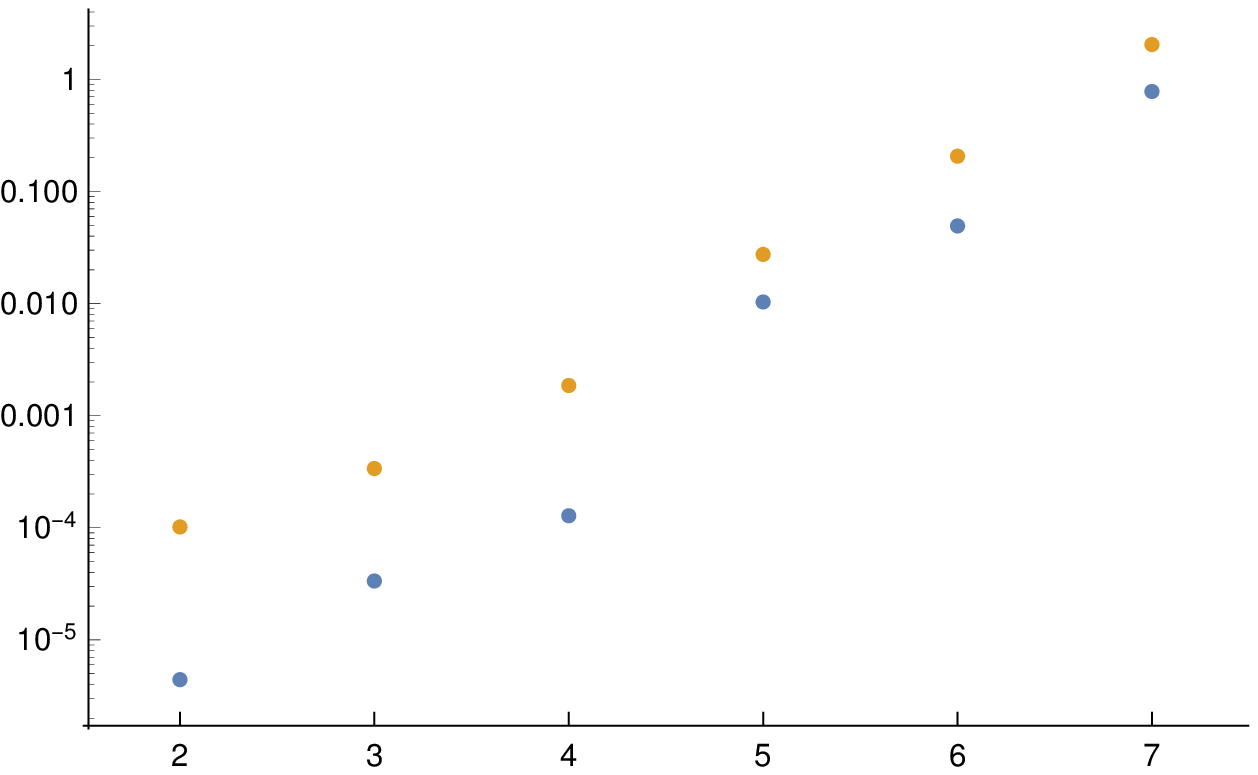}};
    \node at (0,0){$m$};
    \node [rotate=90]at (-6.0,3.5){$t_{ev} / s$};
\end{tikzpicture}
\caption{Average evaluation time of GPLs as a function of the weight
$m$ using \name{} (blue) and GiNaC (yellow). Even though \name{}
remains faster, the lead decreases with increasing weight.}
\label{fig:hi}
\end{figure}

As a last example we considered GPLs of higher weight. While there is
in principle no limitation for the number of parameters that can be
evaluated with the implemented algorithm, in practice the evaluation
can become very slow for weights above $m>7$, depending on the
complexity of the parameters. We have created some more or less
realistic examples for high-weight GPLs by shuffling parameters of the
GPLs appearing in the zeroth set tested above, i.e. the GPLs entering
the heavy-to-light form factor~\cite{Engel:2018fsb}.  The average
evaluation times are plotted in Figure~\ref{fig:hi} as a function of
$m$.

All of these tests were performed on a computer with Intel i5 Kaby
Lake R 1.7GHz processor.

\section{Conclusion}\label{sec:conclusion}
We have presented \name{}, a numerical routine for the fast evaluation
of GPLs. Compared to the current state-of-the-art, \name{} does not
require a framework for symbolic manipulation and is therefore much
faster. GPLs of weight $\le 5$ can now be evaluated fast enough to
allow numerical integration in a Monte Carlo framework.

\section*{Acknowledgement}
We would like to thank Emanuele Bagnaschi, Pulak Banerjee, Tim Engel,
Lukas Fritz, Thomas Gehrmann, Ben Pullin, William J. Torres Bobadilla,
Xiaofeng Xu, Lilin Yang, and Roman Zwicky for comments on the
usability of the code as well as the manuscript.

YU acknowledges support by the Swiss National Science Foundation (SNF)
under contract 200021\_178967.

\clearpage

\bibliographystyle{JHEP-no-et-al}
\bibliography{../../muon_ref}{}

\end{document}